\documentstyle[12pt]{article}

\catcode`@=11
\def\un#1{\relax\ifmmode\@@underline#1\else
        $\@@underline{\hbox{#1}}$\relax\fi}
\catcode`@=12

\let\du=\d                      

\def\a{\alpha}
\def\b{\beta}
\def\c{\chi}
\def\d{\delta}

\def\f{\phi}
\def\g{\gamma}
\def\h{\eta}

\def\j{\psi}

\def\l{\lambda}
\def\m{\mu}

\def\o{\omega}
\def\p{\pi}
\def\q{\theta}
\def\r{\rho}
\def\s{\sigma}

\def\z{\zeta}
\def\D{\Delta}
\def\F{\Phi}

\def\J{\Psi}
\def\L{\Lambda}
\def\O{\Omega}

\def\S{\Sigma}
\def\U{\Upsilon}
\def\X{\Xi}

\def\ve{\varepsilon}

\def\vr{\varrho}

\def\vq{\vartheta}

\def\ce{{\cal E}}

\def\ch{{\cal H}}

\def\cm{{\cal M}}
\def\cn{{\cal N}}

\def\cp{{\cal P}}

\def\ct{{\cal T}}

\def\cv{{\cal V}}

\def\cx{{\cal X}}
\def\cy{{\cal Y}}

\def\bo{{\raise.15ex\hbox{\large$\Box$}}}               
\def\pa{\partial}                                       
\def\de{\nabla}                                         
\def\pr{\prod}                                          
\def\TH{{\raise.2ex\hbox{$\displaystyle \bigodot$}\mskip-4.7mu \llap H \;}}
\def\face{{\raise.2ex\hbox{$\displaystyle \bigodot$}\mskip-2.2mu \llap {$\ddot
        \smile$}}}                                      
\def\dg{\sp\dagger}                                     

\def\sp#1{{}^{#1}}                              
\def\Hat#1{\widehat{#1}}                        
\def\VEV#1{\left\langle #1\right\rangle}        
\def\leftrightarrowfill{$\mathsurround=0pt \mathord\leftarrow \mkern-6mu
        \cleaders\hbox{$\mkern-2mu \mathord- \mkern-2mu$}\hfill
        \mkern-6mu \mathord\rightarrow$}
\def\dvec#1{\vbox{\ialign{##\crcr
        \leftrightarrowfill\crcr\noalign{\kern-1pt\nointerlineskip}
        $\hfil\displaystyle{#1}\hfil$\crcr}}}           
\def\dt#1{{\buildrel {\hbox{\LARGE .}} \over {#1}}}     

\def\frac#1#2{{\textstyle{#1\over\vphantom2\smash{\raise.20ex
        \hbox{$\scriptstyle{#2}$}}}}}                   
\def\ha{\frac12}                                        
\def\sfrac#1#2{{\vphantom1\smash{\lower.5ex\hbox{\small$#1$}}\over
        \vphantom1\smash{\raise.4ex\hbox{\small$#2$}}}} 
\def\bfrac#1#2{{\vphantom1\smash{\lower.5ex\hbox{$#1$}}\over
        \vphantom1\smash{\raise.3ex\hbox{$#2$}}}}       
\def\afrac#1#2{{\vphantom1\smash{\lower.5ex\hbox{$#1$}}\over#2}}    

\def\[{\lfloor{\hskip 0.35pt}\!\!\!\lceil}
\def\]{\rfloor{\hskip 0.35pt}\!\!\!\rceil}

\def\du#1#2{_{#1}{}^{#2}}

\def\fracm#1#2{\hbox{\large{${\frac{{#1}}{{#2}}}$}}}
\def\half{{\fracm12}}
\def\ha{\half}
\def\tr{{\rm tr}}

\def\ula{{\underline a}} \def\ulb{{\underline b}}

\def\un{\underline}
\def\fracmm#1#2{{{#1}\over{#2}}}

\def\low#1{{\raise -3pt\hbox{${\hskip 0.75pt}\!_{#1}$}}}

\def\Dot#1{\buildrel{_{_{\hskip 0.01in}\bullet}}\over{#1}}
\def\dt#1{\Dot{#1}}

\def\Hat#1{\widehat{#1}}

\newskip\humongous \humongous=0pt plus 1000pt minus 1000pt
\def\caja{\mathsurround=0pt}
\def\eqalign#1{\,\vcenter{\openup2\jot \caja
        \ialign{\strut \hfil$\displaystyle{##}$&$
        \displaystyle{{}##}$\hfil\crcr#1\crcr}}\,}
\newif\ifdtup

\def\ref#1{$\sp{#1)}$}

\def\pl#1#2#3{Phys.~Lett.~{\bf {#1}B} (19{#2}) #3}
\def\np#1#2#3{Nucl.~Phys.~{\bf B{#1}} (19{#2}) #3}

\def\pr#1#2#3{Phys.~Rev.~{\bf D{#1}} (19{#2}) #3}
\def\cqg#1#2#3{Class.~and Quantum Grav.~{\bf {#1}} (19{#2}) #3}
\def\cmp#1#2#3{Commun.~Math.~Phys.~{\bf {#1}} (19{#2}) #3}
\def\jmp#1#2#3{J.~Math.~Phys.~{\bf {#1}} (19{#2}) #3}

\def\ijmp#1#2#3{Int.~J.~Mod.~Phys.~{\bf A{#1}} (19{#2}) #3}

\def\ibid#1#2#3{{\it ibid.}~{\bf {#1}} (19{#2}) #3}

\parskip=\medskipamount                 
\thicklines                         

\begin{document}


\thispagestyle{empty}               

\def\border{                                            
        \setlength{\unitlength}{1mm}
        \newcount\xco
        \newcount\yco
        \xco=-24
        \yco=12
        \begin{picture}(140,0)
        \put(-20,11){\tiny Institut f\"ur Theoretische Physik Universit\"at
Hannover~~ Institut f\"ur Theoretische Physik Universit\"at Hannover~~
 Institut f\"ur Theoretische Physik Hannover}
        \put(-20,-241.5){\tiny Institut f\"ur Theoretische Physik Universit\"at
Hannover~~ Institut f\"ur Theoretische Physik Universit\"at Hannover~~
 Institut f\"ur Theoretische Physik Hannover}
        \end{picture}
        \par\vskip-8mm}

\def\headpic{                                           
        \indent
        \setlength{\unitlength}{.8mm}
        \thinlines
        \par
        \begin{picture}(29,16)
        \put(75,16){\line(1,0){4}}
        \put(80,16){\line(1,0){4}}
        \put(85,16){\line(1,0){4}}
        \put(92,16){\line(1,0){4}}

        \put(85,0){\line(1,0){4}}
        \put(89,8){\line(1,0){3}}
        \put(92,0){\line(1,0){4}}

        \put(85,0){\line(0,1){16}}
        \put(96,0){\line(0,1){16}}
        \put(79,0){\line(0,1){16}}
        \put(80,0){\line(0,1){16}}
        \put(89,0){\line(0,1){16}}
        \put(92,0){\line(0,1){16}}
        \put(79,16){\oval(8,32)[bl]}
        \put(80,16){\oval(8,32)[br]}

        \end{picture}
        \par\vskip-6.5mm
        \thicklines}

\border\headpic {\hbox to\hsize{
ITP--UH--18/93 \hfill June 1993}}\par
\vskip2cm
\begin{center}

{\Large\bf $N=2$ Super-Weyl Symmetry, Super-Liouville Theory and
Super-Riemannian Surfaces} \footnote{Supported in part by the 'Deutsche
Forschungsgemeinschaft'}\\
\vglue.2in

Sergei V. Ketov \footnote{On leave of absence from: High
Current Electronics Institute of the Russian Academy of Sciences, Siberian
Branch, Akademichesky~4, Tomsk 634055, Russia} \\
{\it Institut f\"ur Theoretische Physik, Universit\"at Hannover}\\
{\it Appelstra\ss{}e 2, 30167 Hannover, Germany}\\

and\\

Sven-Olaf Moch \footnote{Supported in part by the
'Studienstiftung des deutschen Volkes'}\\
{\it II. Institut f\"ur Theoretische Physik, Universit\"at Hamburg}\\
{\it Luruper Chaussee 149, 22761 Hamburg, Germany}
\end{center}
\vglue.2in
\begin{center}
{\Large\bf Abstract}
\end{center}


The finite form of the $N=2$ super-Weyl transformations in the chiral and
twisted-chiral irreducible formulations of the two-dimensional $N=2$ superfield
supergravity are found in $N=2$ superspace. The super-Weyl anomaly of the
$N=2$ extended fermionic string theory is computed in terms of the $N=2$
superfields, by using a short time expansion of the $N=2$ chiral heat kernel.
The super-Weyl invariant $N=2$ superconformal structure is introduced, and a
new definition of the $N=2$ super-Riemannian surfaces is proposed.
\vglue.2in
{\it Intern. Physics Classification \#'s} ~~0465, 1117, 1130
\vglue.2in
\newpage


\section{Introduction}

Since the work of Polyakov \cite{p81}, much attention has been paid to the
two-dimensional quantum gravity and its supersymmetric extensions. Their
better understanding is crucial for getting more insights into the structure
of (super)-conformal field theories, and critical or non-critical
(super)-string models formulated on the (super)-Riemannian surfaces.

The two-dimensional supergravities can be formulated and investigated either
in components or in superfields. Each approach has its own obvious advantages
and disadvantages, and they are always  complementary to each other.
The $N=1$ supergravity in two dimensions has been investigated in detail, both
in components and in superfields \cite{h79}, and its applications to the
$N=1$ fermionic string theory (also called the NSR model) are well-known
\cite{dp88r}. \footnote{See ref.~\cite{bmg86} for the 'heterotic' case of the
$(1,0)$ supergravity.} As for the $N=2$ or $(2,2)$ two-dimensional
supergravity, most
of its applications (see, e.g., refs.~\cite{ft81,bn86,mm88} for the critical
$N=2$ strings and refs.~\cite{dhk90,abk90} for the non-critical $N=2$ strings)
 have been carried out in components or in the so-called $N=2$ superconformal
gauge, despite of the known formulations of this theory in the full-fledged
$N=2$ curved superspace \cite{hp87,glo89,ghn89,a90}. This seems to be related
 to the fact that some of the relevent elements of the $N=2$ superspace
description of $N=2$ supergravity are yet to be completed. In particular, we
believe it to be certainly true for the {\it finite} form of the relevant
$N=2$ super-Weyl transformations  and a  calculation of the $N=2$
super-Weyl anomaly in terms of the $N=2$ superfields. To the best of our
knowledge, it is apparently missing in the literature. Although the full
$N=2$ superspace description of the $N=2$ supergravity in the $N=2$ (curved)
superspace is highly redundant, knowing the finite form of the $N=2$ super-Weyl
transformations  is important to set up the invariant definition of the $N=2$
super-Riemannian surfaces and the $N=2$ superconformal gauge. It also
matters in  establishing the $N=2$ generalisation of the uniformisation
theorem playing the crucial role in the theory of the (super)-Riemannian
surfaces. The use of the $N=2$ superspace is the best way to uncover the
existence of several different versions of the $N=2$  supergravity
\cite{hp87,glo89,ghn89,a90}, which is very obscure in the component approach.

Our paper is organised as follows. In sect.~2 we formulate the chiral version
of $N=2$ supergravity in superspace and calculate the finite $N=2$ superfield
form of its $N=2$ super-Weyl symmetry transformations. In sect.~3, the similar
results are obatined for the twisted-chiral formulation of the theory. The
$N=2$ fermionic string action in terms of the $N=2$
superfields is discussed in sect.~4. The covariant  $N=2$ superconformal
structure is used to define the $N=2$ super-Riemannian surfaces. The related
definitions of the $N=2$ super-Teichm\"uller and  super-moduli spaces
are given in that section too. Sect.~5 is devoted to the computation of the
$N=2$ super-Weyl anomaly in $N=2$ superspace. It results in the $N=2$
super-Liouville effective theory, as it should. Sect.~6 comprises our
conclusion.
\vglue.2in

\section{Complex Supergeometry in Superspace}

When dealing with spinors on a string world-sheet, one should take into
account the delicate relation which exists between their descriptions on
the Minkowski and Euclidean world-sheets, and the associated spin structure.
The Majorana-Weyl (MW) spinors can only be introduced in Minkowski space, while
defining the super-Riemann surfaces (SRS) is based on the Euclidean
formulation. That's why we find appropriate to start with the $N=2$
supergeometry by using the Minkowski signature, and formally stick to the
Euclidean formulation when introducing the $N=2$ SRS. Keeping in
use both formulations is also important for holomorphic factorisation
\cite{dp88r}. There is in general a topological obstruction to introduce
spinors on a given (Euclidean) manifold $\X$, in order to make possible a
consistent choice of spin structure. Namely, one should have $w_2(\X)=0$,
where $w_2$ is the second Stiefel-Whitney class \cite{a71m,dp86m}. Since the
$w_2(\X)$ vanishes for the {\it oriented} surfaces $\X$, they will be the only
 ones we are going to consider.

Supersymmetric theories can be handled either in components or in superfields.
To construct the correct quantum measure and analyse the anomaly structure of
the $N=2$ superstrings, it is quite appropriate to use the superfield
formalism where the $N=2$ supersymmetry on the world-sheet is manifest.
The natural setting is provided by the $N=2$ (left-right symmetric) curved
superspace of the $(2,2)$ supergravity in two dimensions \cite{hp87,glo89}.
The formal construction of the corresponding $N=2$ supermoduli space and  $N=2$
 SRS goes along the lines of the conventional $N=1$ case \cite{dp88r,adp89},
the important differences being emphasized below. As for discussing global
(or topological) issues, the component approach seems to be more appropriate.

The $N=2$ (flat) superspace coordinates in two dimensions are
$$z^A=(x^a,\q^{\a},\bar{\q}^{\dt{\a}})~,\quad a=0,1,\quad \a=+,-~,\eqno(2.1)$$
 where $\q$'s represent two Grassmannian (anticommuting) complex spinor
coordinates, $(\q^{\a})^{\dg}=\bar{\q}^{\dt{\a}}$. The spinorial covariant
derivatives in the flat $N=2$ superspace are
$$D_{\a}=\pa_{\a}+i\bar{\q}^{\dt{\a}}(\g^a)_{\dt{\a}\a}\pa_a~,$$
$$\bar{D}_{\dt{\a}}=\bar{\pa}_{\dt{\a}}+i\q^{\a}(\g^a)_{\a\dt{\a}}\pa_a~,
\eqno(2.2)$$
and their conjugates, where $\pa_a=\pa/\pa x^a$, $\pa_{\a}=\pa/\pa\q^{\a}$ and
$\bar{\pa}_{\dt{\a}}=\pa/\pa\bar{\q}^{\dt{\a}}$. They satisfy an algebra
$$ \{D_{\a},\bar{D}_{\dt{\a}}\}=2i(\g^a)_{\a\dt{\a}}\pa_a~,\quad
\{D_{\a},D_{\b}\}=\{\bar{D}_{\dt{\a}},\bar{D}_{\dt{\b}}\}=0~.\eqno(2.3)$$

The {\it curved} $N=2$ superspace can be described by the superzweibein
$E\du{M}{A}$ with its inverse $E\du{A}{M}$, the spin superconnection $\o_A$ and
the $U_{\rm V}(1)$ superconnection $\vr_A$. Such formulation is based on the
structure group $SO(1,1)\times U_{\rm V}(1)$ in the tangent space, \footnote{
In two dimensions a general $(p,q)$-supersymmetry algebra may have $p(q)$
left(right)-handed Majorana supersymmetry charges, so that the internal
symmetry group is $SO(p)\times SO(q)$, in general. When $p=q=N=2$, one has
$U_{\rm V}(1)\times U_{\rm A}(1)$.} and leads to the {\it
chiral} (irreducible) version of the $N=2$ supergravity with $4+4$ (off-shell)
components corresponding to a chiral $N=2$ scalar superfield as a compensator
 \cite{glo89,ghn89}. A very natural but naive choice of the $N=2$
superspace with an $SO(1,1)\times U_{\rm V}(1)\times U_{\rm A}(1)$ tangent
space group leads to a {\it reducible} supergravity multiplet with $8+8$
off-shell degrees of freedom comprising a real scalar $N=2$ superfield, as was
 shown in ref.~\cite{hp87}. Taking $SO(1,1)\times U_{\rm A}(1)$ as a structure
 group is another option,  which also results  in the
irreducible $4+4$ off-shell supergravity multiplet, corresponding to the
two-dimensional {\it twisted-chiral} $N=2$ superfield as a compensator
\cite{hp87,ghn89,a90}. However, the anomaly structure of this version of the
$N=2$ superfield supergravity is expected to be more complicated.

Given the superzweibein and superconnections, the curved superspace covariant
derivative can be defined as
$$\de_A=E\du{A}{M}D_M +\o_A\cx + i\vr_A\cy~\equiv E_A +\O_A~,\eqno(2.4)$$
where $D_M$ is the rigid (flat) superspace covariant derivative $D_M=(\pa_m,
\pa_{\m},\bar{\pa}_{\dt{\m}})$ introduced above, $\cx$ and $\cy$ are the
Lorentz and $U_{\rm V}(1)$ symmetry generators, respectively,
$$\eqalign{
\[\cx,O_a\]=\ve\du{a}{b}O_b~, ~~&~ \[\cx,O_{\a}\]=\ha(\g_3)\du{\a}{\b}O_{\b}~,
\cr \[\cy,O_a\]=0~,~~&~ \[\cy,O_{\a}\]=\ha i(\g_3)\du{\a}{\b}O_{\b}~.\cr}
\eqno(2.5)$$

The supertorsion $T\du{AB}{C}$, supercurvature $R_{AB}$ and superfield
strength $F_{AB}$  are defined by the (graded) commutations of the
covariant derivatives,
$$\[\de_A,\de_B\} =T\du{AB}{C}\de_C +R_{AB}\cx +iF_{AB}\cy~,\eqno(2.6)$$
and they are all tensors with respect to the reparametrisations of the $N=2$
superspace coordinates.

The  curved $N=2$ superspace geometry is highly reducible even
off-shell, and it is too general to describe $N=2$ SRS. The off-shell
supergravity in superspace is actually described by some constraints on the
supertorsion, which reduce a number of the independent components to the
minimal one \cite{ggrs}. In this  respect, the two-dimensional $N=2$
supergravity is quite similar to its four-dimensional $N=1$ supergravity
counterpart, where the relevant constraints on the superspace torsion comprise
 the so-called 'conventional' constraints, the '(chiral)
representation-preserving' constraints and the 'conformal' ones \cite{ggrs}.
The latter are just
necessary to reduce the super-Weyl parameter (see below) to the irreducible
$N=2$ superfield  \cite{hp87}. As a result of the constraints and the
subsequent corollaries from the superspace Bianchi identities, all the
torsion, curvature and field-strength tensor superfields can be expressed in
terms of a smaller number of (generically constrained) superfields. As for the
{\it chiral} version of the $N=2$ supergravity theory, \footnote{The different
{\it twisted-chiral} version of the $N=2$ supergravity is considered in the
next section.} the relevant outcome for the (anti)-commutation relations (2.6)
 is given by \cite{glo89,a90}
$$\{\de_{\a},\de_{\b}\}=2(\g_3)_{\a\b}\bar{R}(\cx+i\cy)~,$$
$$\{\de_{\a},\bar{\de}_{\dt{\b}}\}=2i(\g^c)_{\a\dt{\b}}\de_c~,$$
$$\[\de_{\a},\de_b\]=\ha i(\g_b)\du{\a}{\dt{\b}}\left[\bar{R}
\bar{\de}_{\dt{\b}}-(\g_3)\du{\dt{\b}}{\dt{\g}}(\bar{\de}_{\dt{\g}}\bar{R})
(\cx+i\cy)\right]~,$$
$$\[\de_a,\de_b\]={\frac 14}\ve_{ab}\left\{2(\g_3)^{\a\b}(\de_{\a}R)\de_{\b}
+ 2(\g_3)^{\dt{\a}\dt{\b}}(\bar{\de}_{\dt{\a}}\bar{R})\bar{\de}_{\dt{\b}}
\right.$$
$$\left. + \left[\de^2R+\bar{\de}^2\bar{R}-4R\bar{R}\right]\cx-
i\left[\de^2R - \bar{\de}^2\bar{R}\right]\cy\right\}~.\eqno(2.7)$$
In eq.~(2.7) all the tensor-component $N=2$ superfields of the torsion,
curvature and field-strength  are expressed in terms of a single scalar
$N=2$ complex superfield $R$.  The first line of eq.~(2.7) implies that this
superfield is (covariantly) chiral,
$$\bar{\de}_{\dt{\a}}R=\de_{\a}\bar{R}=0~.\eqno(2.8)$$
Having imposed the supertorsion constraints, one can express, as usual, the
superconnections in terms of the superzweibein. For instance, the constraint
$T\du{\a\b}{\dt{\g}}=E\du{\b}{M}E\du{\a}{N}\left(\de_N\bar{E}\du{M}{\dt{\g}}
-\de_M\bar{E}\du{N}{\dt{\g}}\right)=0$ implies
$$(\o_M\pm\vr_M)=(\g_3\pm 1)\du{\dt{\g}}{\dt{\b}}\bar{E}\du{\dt{\b}}{N}\left(
D_N\bar{E}\du{M}{\dt{\g}}- D_M\bar{E}\du{N}{\dt{\g}}\right)~.\eqno(2.9)$$

The complex supergeometry of the $N=2$ supergravity is invariant under those
transformations which preserve the constraints. One usually finds convenient
to look at the infinitesimal variations first, in the form $H\du{A}{B}
=E\du{A}{M}\d E\du{M}{B}$ and $\f\du{A,B}{C}=E\du{A}{M}\d \O\du{M,B}{C}$, where
$\O$ is the total connection, $\O_A=\o_A\cx + i \vr_A\cy$. The corresponding
variation of the supertorsion components reads \cite{hp87}
$$\d T\du{AB}{C}=\de_A H\du{B}{C}-(-1)^{AB}\de_B H\du{A}{C}
+T\du{AB}{D}H\du{D}{C}$$
$$-H\du{A}{D}T\du{DB}{C}+(-1)^{AB}H\du{B}{D}T\du{DA}{C}+\f\du{A,B}{C}-(-1)^{AB}
\f\du{B,A}{C}~.\eqno(2.10)$$
Since the supertorsion constraints, not all of the $H$'s are actually
independent. The $H\du{\a}{b}$, $H\equiv H\du{\a}{\a}$, $(\g_3H)\equiv
(\g_3)\du{\a}{\b}H\du{\b}{\a}$ and their complex conjugates can be chosen to
represent a complete set of the independent ones.

By construction, the super-reparametrisational, super-Lorentz and super-phase
$U_{\rm V}(1)$ local transformations in the $N=2$ superspace  are
always among the symmetries of the theory, and we are not going to discuss
them in any detail (see, however, refs.~\cite{hp87,a90}). Instead, we want to
concentrate on another local symmetry which is closely related to  two
bosonic dimensions we are working in, and is crucial for the $N=2$ fermionic
strings, namely the $N=2$ {\it super-Weyl} invariance. Taken together,
the local symmetries are enough to gauge away all the $N=2$ supergravity
fields in each given coordinate patch, but not globally. The fact that the
complex supergeometry of the $N=2$ supergravity is {\it superconformally flat}
was first noticed by Howe and Papadopoulos \cite{hp87}, but, to the best of
our knowledge, the {\it finite} $N=2$ super-Weyl transformations in the
irreducible formulations of the $N=2$ supergravity in $N=2$ superspace were
not calculated.

In their infinitesimal form, the $N=2$ super-Weyl transformations are
\cite{hp87}
$$H\du{a}{b}=\d\du{a}{b}(\l+\bar{\l})~,\quad H\du{\a}{\b}=\d\du{\a}{\b}
\bar{\l}~,\quad H\du{a}{\dt{\b}}=\fracmm{i}{2}(\g_a)^{\dt{\b}\g}\de_{\g}\l~,
\eqno(2.11)$$
where $\l$ is an infinitesimal {\it chiral} $N=2$ superfield parameter. It
follows from eq.~(2.10) that the symmetry transformations (2.11) are the
invariance of the constraints (2.7).

It is convenient to use $N=2$ super-differential forms here, in terms of which
 the superzweibein takes the form
$E^A=dz^M E\du{M}{A}=(E^a,E^{\a},\bar{E}^{\dt{\a}})$,
and similarly for the derivatives. The supertorsion and the superspace Bianchi
identities can then be conveniently represented as \cite{hp87}
$$T^C=\ha E^B E^A T\du{AB}{C}=DE^C +E^B\O\du{B}{C}\equiv\de E^A~,\eqno(2.12)$$
and
$$\de T^A=E^BR\du{B}{A}~,\quad DF=0~,\eqno(2.13)$$
respectively, where $R\du{A}{B}=D\O\du{A}{B}+\O\du{A}{C}\O\du{C}{B}$,
$R\du{a}{b}=\ve\du{a}{b}F$ and $R\du{\a}{\b}=\ha(\g_3)\du{\a}{\b}F$.

The form of eq.~(2.11) suggests the following ansatz for the finite $N=2$
super-Weyl transformations with the chiral $N=2$ superfield parameter $\L$
$$\hat{E}^a=SE^a~,~~{\rm where}~~ S\equiv \L\bar{\L}~,$$
$$\hat{E}^{\a}=\bar{\L}E^{\a} + E^c(\g_c)^{\a\dt{\b}}\bar{\r}_{\dt{\b}}~,$$
$$\hat{\bar{E}}^{\dt{\a}}=\L\hat{E}^{\dt{\a}} + E^c(\g_c)^{\dt{\a}\b}\r_{\b}~,
\eqno(2.14)$$
which some superfields $\r_{\b},\bar{\r}_{\dt{\b}}$ to be determined by
evaluating the supertorsion components $\hat{T}^c$ in the two different ways.
First, their definition according to eq.~(2.12) in terms of the $\hat{E}^A$
and the (yet unknown) $\hat{\o}$ and $\hat{\vr}$ implies
$$\hat{T}^c=\ha \hat{E}^B\hat{E}^A T\du{AB}{c}=ST^c+S\left[(\hat{\o}-\o)\cx +
i(\hat{\vr}-\vr)\cy\right]E^c+(\de S)E^c$$
$$=iS\bar{E}^{\dt{\b}}E^{\a}(\g^c)_{\a\dt{\b}}+SE^b\ve\du{b}{c}\U
+(\de S)E^c~,\eqno(2.15a)$$
where $\left[(\hat{\o}-\o)\cx +i(\hat{\vr}-\vr)\cy\right]E^c\equiv
E^b\ve\du{b}{c}\U$. Second, the ansatz (2.14) yields another equation for
the same tensor $\hat{T}^c$, and it is consistent with eq.~(2.15a) since it
appears to have the same structure, namely
$$\hat{T}^c=i(\g^c)_{\a\dt{\b}}\left[ S\bar{E}^{\dt{\b}}E^{\a} +
\L\bar{E}^{\dt{\b}}E^a(\g_a)^{\a\dt{\d}}\bar{\r}_{\dt{\d}} +
\bar{\L}E^{\a}E^a(\g_a)^{\dt{\b}\g}\r_{\g}\right]~.\eqno(2.15b)$$
Comparing the coefficients at $E^c$ and $E^d\ve\du{d}{c}$ in eq.~(2.15) gives
rise to the  equations
$$\de S=i\L\bar{E}^{\dt{\b}}\bar{\r}_{\dt{\b}}+i\bar{\L}E^{\a}\r_{\a}~,$$
$$S\U=-i\L\bar{E}^{\dt{\b}}(\g_3\bar{\r})_{\dt{\b}}-
i\bar{\L}E^{\a}(\g_3\r)_{\a}~.\eqno(2.16)$$
The first line of eq.~(2.16) fixes the $\r_{\a}$ and $\bar{\r}_{\dt{\a}}$.
Hence, the $N=2$ super-Weyl transformation of the superzweibein is given by
$$\hat{E}^a=SE^a~,\qquad S=\L\bar{\L}~,$$
$$\hat{E}^{\a}=\bar{\L}E^{\a} -iE^b(\g_b)^{\a\dt{\b}}\bar{\de}_{\dt{\b}}
\bar{\L}~,$$
$$\Hat{\bar{E}}{}^{\dt{\a}}=\L\bar{E}^{\dt{\a}}-i E^b(\g_b)^{\dt{\a}\b}\de_{\b}
\L~.\eqno(2.17)$$
Since $E\du{A}{M}E\du{M}{B}=\d\du{A}{B}$ and similarly for the superzweibein
components with hats, it follows from eq.~(2.17) that the inverse
superzweibein transforms as
$$\hat{E}_a=S^{-1}E_a + iS^{-1}\bar{\L}^{-1}(\g_a)^{\b\dt{\g}}
(\bar{\de}_{\dt{\g}}\bar{\L})E_{\b} + iS^{-1}\L^{-1}(\g_a)^{\dt{\b}\g}
(\de_{\g}\L)\bar{E}_{\dt{\b}}~,$$
$$\hat{E}_{\a}=\bar{\L}^{-1}E_{\a} ~,$$
$$\hat{\bar{E}}_{\dt{\a}}=\L^{-1}\bar{E}_{\dt{\a}} ~.\eqno(2.18)$$
Notably, the $N=2$ super-Weyl transformation of the spinor components of the
inverse superzweibein in eq.~(2.18) is very simple.

The super-Weyl transformation of the spinor components of the Lorentz
superconnection follows from the second line of eq.~(2.16), and it takes the
form
$$\hat{\o}_{\a}=\bar{\L}^{-1}\o_{\a}-\bar{\L}^{-1}\L^{-1}(\g_3)\du{\a}{\g}
\de_{\g}\L~.\eqno(2.19)$$
We have checked that the same result also follows by exploiting the explicit
form of dependence of the Lorentz superconnection upon the superzweibein,
i.e. by using  eqs.~(2.9), (2.17) and (2.18).  Similarly, one finds
$$\hat{\vr}_{\a}=\bar{\L}^{-1}\vr_{\a} - \bar{\L}^{-1}\L^{-1}(\g_3)
\du{\a}{\g}\de_{\g}\L~.\eqno(2.20)$$
The super-Weyl transformations of the vector components of the superconnections
are more complicated, but, fortunately, we don't need them in what follows.

The $N=2$ super-Weyl transformations of the spinorial covariant derivatives in
the $N=2$ superspace are straightforward to calculate, since they are the
direct corollaries of the transformation rules given above. We find
$$\hat{\de}_{\a}=\bar{\L}^{-1}\de_{\a}-\bar{\L}^{-1}\L^{-1}(\g_3)\du{\a}{\g}
(\de_{\g}\L)(\cx +i\cy)~.\eqno(2.21)$$
The first line of eq.~(2.7) then implies the transformation law for the
anti-chiral superfield $\bar{R}$ in the form
$$\Hat{\bar{R}}=\bar{\L}^{-2}\left[\bar{R}-2\L^{-1}\de^2\L+2\L^{-2}(\de^{\a}\L)
(\de_{\a}\L)\right]~,\eqno(2.22a)$$
or, equivalently,
$$\Hat{\bar{R}}=e^{-2\bar{\S}}\left(\bar{R}-2\de^2\S\right)~,\quad
\Hat{R}=e^{-2\S}\left( R -2\bar{\de}^2\bar{\S}\right)~,\eqno(2.22b)$$
where the chiral $N=2$ superfield $\S$ has been introduced, $\L\equiv\exp(\S)$.
Eqs.~(2.18) and (2.22) comprise the main results of this section. It should
also be noticed that the superdeterminant \footnote{As for the definition of
the superdeterminant, see eqs.~(A.9) and (A.10) of the Appendix.} $E=\equiv
{\rm sdet}(E\du{M}{A})$ is $N=2$ super-Weyl invariant, $\hat{E}=E$ or $\d E
=0$, which is easily verified. Finally, as for the $N=2$ chiral density $\ce$
to be defined as $\ce\equiv -\ha\left(\bar{\de}^2-4R\right)E$, we get
$$\hat{\ce}=e^{-2\S}\left[\ce-4(\bar{\de}^2\bar{\S})E\right]~,\quad
\ce=e^{2\S}\hat{\ce} +4(\bar{\de}^2\bar{\S})\hat{E}~.\eqno(2.23)$$
\vglue.2in

\section{Twisted $N=2$ Superfield Supergravity}

The different set of the two-dimensional $N=2$ supergravity constraints in
$N=2$ superspace can be obtained by truncating the four-dimensional $N=1$
superfield supergravity down to two dimensions.  The alternative set of the
two-dimensional $N=2$ supergravity constraints takes the form \cite{glo89,a90}
$$\{\de_{\a},\de_{\b}\}=0~,$$
$$\{\de_{\a},\bar{\de}_{\dt{\b}}\}=i(\g^c)_{\a\dt{\b}}\de_c +\left(
iC_{\a\dt{\b}}H+(\g_3)_{\a\dt{\b}}G\right)\cx -\left(C_{\a\dt{\b}}G
+i(\g_3)_{\a\dt{\b}}H\right)\cy_t~,$$
$$\[\de_{\a},\de_b\]=\ha\left[ i(\g_b)\du{\a}{\b}G -(\g_3\g_b)\du{\a}{\b}H
\right]\de_{\b} + i(\g_3\g_b)\du{\a}{\b}(\de_{\b}G)\cx-i(\g_b)\du{\a}{\b}
(\de_{\b}G)\cy_t~,$$
$$\[\de_a,\de_b\]=-\ve_{ab}\left[(\bar{\de}^{\dt{\g}}G)(\g_3)\du{\dt{\g}}{\l}
\de_{\l}+(\de^{\l}G)(\g_3)\du{\l}{\dt{\g}}\bar{\de}_{\dt{\g}}\right.$$
$$\left. -\left(C^{\a\dt{\b}}\de_{\a}\bar{\de}_{\dt{\b}}G-G^2-H^2\right)\cx
-(\g_3)^{\a\dt{\b}}\left(\de_{\a}\bar{\de}_{\dt{\b}}G\right)\cy_t\right]~,
\eqno(3.1)$$
where the new $U(1)$ generator $\cy_t$ has been introduced,
$$\[\cy,O_a\]=0~, \quad  \[\cy,O_{\a}\]=\ha iO_{\a}~.\eqno(3.2)$$
In this new $N=2$ supergravity theory all the supertorsion, supercurvature and
superconnection
components depend on the two real scalar superfields $G$ and $H$, which can be
combined into the single complex $N=2$ superfield to be equivalent to a
twisted chiral $N=2$ scalar superfield \cite{ghr84}. It should be noticed that
 both
(chiral and twisted-chiral) versions of $N=2$ supergravity are derivable by
truncating the off-shell formulation of the two-dimensional $N=4$ supergravity
theory of ref.~\cite{glo89}.

The infinitesimal $N=2$ super-Weyl transformations in the twisted superfield
formulation of the $N=2$ supergravity theory take the form
$$H\du{a}{b}=\d\du{a}{b}\left(\L +\bar{\L}\right)~,$$
$$H\du{\a}{\b}=\ha\left(\d\du{\a}{\b}-(\g_3)\du{\a}{\b}\right)\L +
\ha\left(\d\du{\a}{\b}+(\g_3)\du{\a}{\b}\right)\bar{\L}~,$$
$$H\du{a}{\dt{\b}}=i(\g_a)^{\dt{\d}\g}\de_{\g}\left[\ha\left(
\d\du{\dt{\d}}{\dt{\b}}+(\g_3)\du{\dt{\d}}{\dt{\b}}\right)\L +
\ha\left(\d\du{\dt{\d}}{\dt{\b}}-(\g_3)\du{\dt{\d}}{\dt{\b}}\right)
\bar{\L}\right]~,\eqno(3.3)$$
which is different from that given in ref.~\cite{hp87}. We have checked that
the supertorsion constraints in eq.~(3.1) are invariant with respect to our
$N=2$ super-Weyl transformation laws in eq.~(3.3). The $N=2$ super-Weyl
parameter
in eq.~(3.3) is supposed to be a twisted-chiral superfield, $\ha(1+\g_3)
\du{\a}{\b}\de_{\b}\L=\ha(1-\g_3)\du{\dt{\a}}{\dt{\b}}\bar{\de}_{\dt{\b}}
\bar{\L}=0$.

A derivation of the finite form of the $N=2$ super-Weyl transformations
follows the lines of the chiral case already considered in the previous
section. It results in
$$\hat{E}\du{M}{a}=E\du{M}{a}S~,~~{\rm where}~~ S\equiv \L\bar{\L}~,
\eqno(3.4)$$
$$\hat{E}\du{M}{\a}=\ha\left(E\du{M}{\b}-2iE\du{M}{a}(\g_a)^{\b\dt{\g}}S^{-1}
\bar{\de}_{\dt{\g}}S\right)\left[\left(\d\du{\b}{\a}-(\g_3)\du{\b}{\a}
\right)\L +\left(\d\du{\b}{\a}+(\g_3)\du{\b}{\a}\right)\bar{\L}\right]~,$$
$$\hat{\bar{E}}\du{M}{\dt{\a}}=\ha\left(\bar{E}\du{M}{\dt{\b}}-2iE\du{M}{a}
(\g_a)^{\dt{\b}\g}S^{-1}\de_{\g}S\right)\left[\left(\d\du{\dt{\b}}{\dt{\a}}
+(\g_3)\du{\dt{\b}}{\dt{\a}}\right)\L +\left(\d\du{\dt{\b}}{\dt{\a}}
-(\g_3)\du{\dt{\b}}{\dt{\a}}\right)\bar{\L}\right]~,$$
where $M$ is a curved superspace index, $M=(m,\m,\dt{\m})$. Eq.~(3.4) implies
the inverse superzweibein to transform as
$$\hat{E}\du{a}{M}=S^{-1}\left[E\du{a}{M}-2iS^{-1}(\g_a)^{\d\dt{\g}}
(\bar{\de}_{\dt{\g}}S)E\du{\d}{M}-2iS^{-1}(\g_a)^{\dt{\d}\g}(\de_{\g}S)
\bar{E}\du{\dt{\d}}{M}\right]~,$$
$$\hat{E}\du{\a}{M}=\ha\left[\left(\d\du{\a}{\b}-(\g_3)\du{\a}{\b}\right)
\L^{-1} +\left(\d\du{\a}{\b}+(\g_3)\du{\a}{\b}\right)\bar{\L}^{-1}\right]
E\du{\b}{M}~,$$
$$\hat{\bar{E}}\du{\dt{\a}}{M}=\ha\left[\left(\d\du{\dt{\a}}{\dt{\b}}+(\g_3)
\du{\dt{\a}}{\dt{\b}}\right)\L^{-1} +\left(\d\du{\dt{\a}}{\dt{\b}}-(\g_3)
\du{\dt{\a}}{\dt{\b}}\right)\bar{\L}^{-1}\right]\bar{E}\du{\dt{\b}}{M}~.
\eqno(3.5)$$

The super-Weyl transformation rules for the fermionic parts of the
superconnections,
$$\o_M E\du{\a}{M}\equiv\bar{E}\du{\dt{\g}}{N}(\g_3)\du{\dt{\b}}{\dt{\g}}
\left(D_N\bar{E}\du{M}{\dt{\b}}-D_M\bar{E}\du{N}{\dt{\b}}\right)E\du{\a}{M}~,$$
$$\vr_M E\du{\a}{M}\equiv\bar{E}\du{\dt{\b}}{N}\left(D_N\bar{E}\du{M}{\dt{\b}}
-D_M\bar{E}\du{N}{\dt{\b}}\right)E\du{\a}{M}~,\eqno(3.6)$$
 are now straightforward to calculate. We find
$$\hat{\o}_{\a}=\ha\left[\left(\d\du{\a}{\b}-(\g_3)\du{\a}{\b}\right)\L^{-1} +
\left(\d\du{\a}{\b}+(\g_3)\du{\a}{\b}\right)\bar{\L}^{-1}\right]\left[
\o_{\b}-4S^{-1}(\g_3)\du{\b}{\g}(\de_{\g}S)\right]~,$$
$$\hat{\vr}_{\a}=\ha\left[\left(\d\du{\a}{\b}-(\g_3)\du{\a}{\b}\right)\L^{-1} +
\left(\d\du{\a}{\b}+(\g_3)\du{\a}{\b}\right)\bar{\L}^{-1}\right]\left[
\vr_{\b}+4S^{-1}(\de_{\b}S)\right]~.\eqno(3.7)$$
Eq.~(3.7) fixes the Weyl transformations of the superspace covariant
derivatives to the form
$$\hat{\de}_{\a}=\ha\left[\left(\d\du{\a}{\b}-(\g_3)\du{\a}{\b}\right)\L^{-1} +
\left(\d\du{\a}{\b}+(\g_3)\du{\a}{\b}\right)\bar{\L}^{-1}\right]$$
$$\times \left[
\de_{\b}-4S^{-1}\left\{(\g_3)\du{\b}{\g}(\de_{\g}S)\cx-i(\de_{\b}S)\cy
\right\}\right]~,$$
$$\hat{\bar{\de}}_{\dt{\a}}=\ha\left[\left(\d\du{\dt{\a}}{\dt{\b}}+(\g_3)
\du{\dt{\a}}{\dt{\b}}\right)\L^{-1} +
\left(\d\du{\dt{\a}}{\dt{\b}}-(\g_3)\du{\dt{\a}}{\dt{\b}}\right)\bar{\L}^{-1}
\right]$$
$$\times \left[\bar{\de}_{\dt{\b}}-4S^{-1}\left\{(\g_3)\du{\dt{\b}}{\dt{\g}}
(\bar{\de}_{\dt{\g}}S)\cx+i(\bar{\de}_{\dt{\b}}S)\cy\right\}\right]~.
\eqno(3.8)$$

The transformation rules of the superfields $G$ and $H$  follow from the
second line of eq.~(3.1) to be contracted with the charge conjugation
matrix $C^{\a\dt{\b}}$, and eq.~(3.8). The finite $N=2$ super-Weyl
transformations laws turn out to be surprisingly simple, namely
\footnote{Eq.~(3.1) and the obvious identities $\tr(\g_3\g^a)=\tr(\g^a)=0$
have been used to derive eq.~(3.9).}
$$\hat{H}=e^{-\S}e^{-\bar{\S}}\left\{H + 2\de_{\a}(\g_3)^{\a\dt{\b}}
\bar{\de}_{\dt{\b}}(\S+\bar{\S})\right\}~,$$
$$\hat{G}=e^{-\S}e^{-\bar{\S}}\left\{G + 2i\de_{\a}C^{\a\dt{\b}}
\bar{\de}_{\dt{\b}}(\S+\bar{\S})\right\}~,\eqno(3.9)$$
where the twisted chiral superfield $\S$ has been introduced, \footnote{We
use  the notation similar to that of the previous section, since
the chiral and twisted chiral formulations of the $N=2$ supergravity can never
simultaneously appear in one theory.}
$$\L\equiv e^{\S}~,\quad \bar{\L}\equiv e^{\bar{\S}}~.\eqno(3.10)$$
Eq.~(3.9) represents the main result of this section.
\vglue.2in

\section{Superconformal Gauge and $N=2$ SRS}

Eqs.~(2.22) and (3.9) play the crucial role in the complex $N=2$
supergeometry. First, they mean that a curved $N=2$ superspace of the $N=2$
supergravity is superconformally flat, namely  the relevant superfields of the
$N=2$ supergravity (the chiral $(R)$ or the twisted chiral $(G+iH)$ $N=2$
superfield) can be made constants by the $N=2$ super-Weyl transformations.

The component gauge fields of the $N=2$ supergravity are naturally defined via
the expansion of the $N=2$ superspace vector covariant derivative,
$$\de_a|=e^m_a\pa_m + \j_a^{\m}\pa_{\m}
+\bar{\j}_a^{\dt{\m}}\bar{\pa}_{\dt{\m}} +\o_a\cx + iA_a\cy~,\eqno(4.1)$$
where $|$ denotes the $\q=0$ projection. In eq.~(4.1), $e^m_a(x)$ is the
two-dimensional zweibein, $\j_a^{\m}(x)$ is the two-dimensional complex
'gravitino' field, and $A_a(x)$ is an Abelian gauge field. In the Wess-Zumino
gauge, where most of the auxiliary fields required by $N=2$ superfields are
eliminated, the off-shell field contents of the $N=2$ supergravity in
components are $(e^a_m,\j_a^{\m},A_a,R|)$ in the chiral formulation, and
$(e^a_m,\j_a^{\m},A_a,G|,H|)$ in the twisted chiral one.

The $N=2$ superconformal flatness is due to the existence of the (unique)
solution to the $N=2$ chiral Liouville equation for the finite $N=2$
super-Weyl parameters $\S$ and $\bar{\S}$,
$$2\bar{\de}^2\bar{\S}+e^{2\S}\Hat{R}=R~,\eqno(4.2)$$
where $\Hat{R}=const$. The value of the complex constant is constrained by
topology, since one has (in the Wess-Zumino gauge) the relation
$$\fracmm{1}{\p}\int d^2xd^2\q\,\ce R=\fracmm{1}{2\p}\int d^2x\,eR^{(2)}
+\fracmm{i}{2\p}\int d^2x\,eF=\c(\s)+ic~,\eqno(4.3)$$
where the two-dimensional scalar curvature $R^{(2)}(e)$ and the Abelian
field strength $F(A)$, as well as the corresponding Euler characteristic
$\c(\S)=2-2h$ and the first Chern class $c$, have been introduced,
$h,c\in{\bf Z},~h\geq 0$.  The $\ce$ in
eq.~(4.3) means the chiral density, $\ce\equiv-\ha(\bar{\de}^2-4R)E$,
$E\equiv {\rm sdet}(E\du{M}{A})$ and $\bar{\de}_{\dt{\a}}\ce=0$.

Clearly, the $N=2$ {\it flat} superspace is characterized by
$\hat{R}_{\rm flat}=0$ and $\de_{\rm flat}=D$, so that
$$R_{\rm flat}=2\bar{D}^2\bar{\S}~. \eqno(4.4)$$
There are, of course, the topological obstructions (moduli!) when $h>0$. In
addition, eq.~(4.4) is only valid in classical theory, because the $N=2$
super-Weyl invariance is anomalous after quantisation. Eq.~(2.22) also implies
that the superspace constraints of the $N=2$ supergravity can be locally
solved by setting the superzweibein to be equal to the $N=2$
super-Weyl-transformed {\it flat} superzweibein. Such choice constitutes the
$N=2$ {\it superconformal gauge}. This gauge is very convenient for quantum
calculations, just like the conformal gauge is, since the redundant
(super)gravity fields disappear in that gauge, being fixed by the
non-anomalous local symmetries.

The $N=2$ fermionic string (Polyakov-type) action on the $N=2$ supersymmetric
(curved) 'world-sheet' $\X$ is written in terms of the $N=2$ (covariantly)
chiral superfields
$X^{\ula}$, $\ula=0,1,\ldots,d-1$, $\bar{\de}_{\dt{\a}}X=\de_{\a}\bar{X}=0$,
as \footnote{For definiteness, we use the $N=2$ chiral
superfields to represent the $N=2$ scalar matter, or the $N=2$ superstring
coordinates. They could equally be represented by the $N=2$ twisted chiral
superfields.}
$$S_0=\int d^2x d^2\q d^2\bar{\q}\,E \bar{X}^{\ula}X^{\ulb}\h_{\ula\ulb}~,
\eqno(4.5)$$
with some flat 'space-time' metric $\h_{\ula\ulb}$. The most general
renormalisable $N=2$ fermionic string action $S_{\rm str}$ also
includes the 'topological' term of eq.~(4.3) and the 'cosmological' term
$$S_{\rm c}=\m_0\int d^2x d^2\q\, \ce + {\rm h.c.}~,\eqno(4.6)$$
where $\m_0$ is a constant.

Going along the lines of the conventional bosonic and $N=1$ supersymmetric
cases \cite{gsw}, we now perform the Wick rotation and switch to the Euclidean
formulation of the $N=2$ fermionic string theory (4.5) characterized by the
partition function $Z=\sum_{h,c} Z_{h,c}$, where
$$Z_{h,c}=\int \[dE\du{M}{A}\]\[d\O_N\]\d(T)\[d\bar{X}\]\[dX\]
e^{-S_{\rm str}[E,\O,X]}~.\eqno(4.7)$$
In eq.~(4.7) the delta-function symbolizes the $N=2$ supergravity constraints,
 which effectively remove the integration over the superconnection, in
particular.  The
functional integration measure is determined by the generalised ultra-locality
 principle \cite{p86u} and the $N=2$ super-reparametrisational invariance, and
it is based on the following definitions of the norm for the superfields $X$
and $E\du{M}{A}$,
$$\left|\left|\d X^{\ula}\right|\right|^2=\int d^2x d^2\q d^2\bar{\q}\,E
\d X^{\ula}\d X_{\ula}~,$$
$$\left|\left|\d E\du{M}{A}\right|\right|^2=\int d^2x d^2\q d^2\bar{\q}\,E
\left[ \ve^{\a\b}H^a_{\a}H^a_{\b}+c_1H^2 + c_2(\g_3H)^2\right] + {\rm h.c.}~,
\eqno(4.8)$$
where some arbitrary constants $c_1$ and $c_2$ have been introduced. Since
both norms in eq.~(4.8) fail to be $N=2$
super-Weyl invariant, this symmetry is expected to be anomalous, which is the
case when $d\neq 2$ indeed, as is well-known from the component considerations
\cite{ft81,k93}. The detailed form of the $N=2$ super-Weyl anomaly in the $N=2$
superspace will be determined in the next section. To the end of this section,
we are working in the critical dimension $d=2$, where there is no super-Weyl
anomaly.

It is quite natural to assume that the bosonic part ('body') of our
supermanifold (or supersurface) $\X$ forms a Riemann surface $\X|$. We can
then introduce the $N=2$ almost supercomplex structure on $\X$ as follows
$$J\du{M}{N}=E\du{M}{a}\ve\du{a}{b}E\du{b}{N}
+iE\du{M}{\a}(\g_3)\du{\a}{\b}E\du{\b}{N}
+iE\du{M}{\dt{\a}}(\g_3)\du{\dt{\a}}{\dt{\b}}E\du{\dt{\b}}{N}~,\eqno(4.9)$$
which satisfies
$$J\du{M}{N}J\du{N}{P}=-\d\du{M}{P}~,\eqno(4.10)$$
and is {\it invariant} under the $N=2$ super-Weyl transformations, as we
explicitly verified. Similarly to the conventional $N=1$ supersymmetric case
\cite{dp88r}, it is not difficult to show that the $N=2$ almost supercomplex
structure defined above is integrable, and, in particular, this allows us to
globally define $N=2$ superholomorphic coordinates. It can actually be done by
introducing the 1-(super)form $\z^M=dz^M -idz^MJ\du{N}{M}$, having only two
independent components because of eq.~(4.10), and checking that
$d\z^M=0\,({\rm mod}\;\z^N)$ indeed. It happens to be the case just because
 of the $N=2$ supergravity constraints. Hence, we are in a position to define
$N=2$ super-holomorphic and super-antiholomorphic functions $\F$ and
$\bar{\F}$ as solutions to the equations
$$J\du{M}{N}\de_N\F=i\de_M\F~,\quad J\du{M}{N}\de_N\bar{\F}=-i\de_M\bar{\F}~.
\eqno(4.11a)$$
or, equivalently,
$$\de_-\F=0~,\qquad \de_+\bar{\F}=0~,\eqno(4.11b)$$
where $\de_{\pm}$ are the corresponding 'chiral' (with respect to the $N=2$
supercomplex structure $J$) covariant derivatives. The $N=2$ superspace
coordinates associated to the $N=2$ supercomplex structure are termed
the $N=2$ superconformal coordinates. It is now natural {\it to define the}
$N=2$ {\it SRS as an} $N=2$ {\it supersurface equipped with an} $N=2$ {\it
supercomplex structure}, in a complete analogy with the $N=0$ and $N=1$ cases
\cite{dp88r}. \footnote{We always assume here that all the supermanifolds we
consider are the
supermanifolds in the conventional sense \cite{dw83}, with all the
non-trivial topology due to the bosonic 'body' only.} On the
$N=2$ SRS, the (local) coordinate patches should exist, whose transition
functions (instructing how to put those patches together) are superholomorphic.
This would establish the contact with the alternative (presumably, equivalent)
description of the $N=2$ SRS introduced earlier \cite{c87,m88} as the
$1|2$-(complex)dimensional superconformal manifolds with $N=2$ superconformal
transition functions in overlapping regions. \footnote{The supercoordinate
transformation is called superconformal provided the flat supercovariant
derivative $D_+$ transforms as a superconformal tensor.}

The local symmetries of the theory (4.5) in the curved $N=2$ superspace
comprise (i) $N=2$ super-diffeomorphisms ${\rm 2sDiff}(\X)$, (ii) $N=2$
supersymmetric Lorentz transformations ${\rm 2sL}$, (iii)  $N=2$
supersymmetric Abelian (phase) transformations ${\rm 2sU(1)}$, and (iv) $N=2$
super-Weyl transformations ${\rm 2sWeyl}(\X)$. Let $\d V_A(z)$, $\d L(z)$,
$\d M(z)$ and $\d\S(z)$ be the corresponding infinitesimal $N=2$ superfield
local parameters, respectively. Using the symmetries (iii) and (iv), the
$H$ and $(\g_3H)$ can be eliminated from eq.~(4.8) without topological
obstructions, which explains the redundancy of the coefficients $c_1$ and
$c_2$ in this equation. The variation $H^a_{\a}$ under the infinitesimal $N=2$
super-diffeomorphisms, {\it orthogonal} to the action of the $N=2$ super-Weyl
symmetry, is governed by the $N=2$ super-differential operator $\cp_1$ of the
form
$$\left(\cp_1\d V\right)\du{\a}{b}=\ha\left(\g^c\g^b\right)\du{\a}{\b}\de_{\b}
\d V_c~,\eqno(4.12)$$
in a complete analogy with the bosonic and $N=1$ supersymmetric cases
\cite{dp88r}, where the two-dimensional identity $\g_a\g^b\g^a=0$ has been
used.

We are actually interested in the different $N=2$ supergeometries which are
not related by the $N=2$ super-diffeomorphisms, $N=2$ local
Lorentz, $U(1)$ or Weyl transformations. So, let's consider an arbitrary total
variation $\{H\du{A}{B}\}$, which  can be decomposed as
$$\{H\du{A}{B}\}=\{\d\S\}\oplus\{\d L\}\oplus\{\d M\}\oplus\{{\rm Range}~
\cp_1\}\oplus\{{\rm Ker}\cp_1^{\dg}\}~.\eqno(4.13)$$
The elements of ${\rm Ker}\cp_1$ are natural to term  the $N=2$ superconformal
Killing vectors, while the elements of ${\rm Ker}\cp_1^{\dg}$ should be termed
the $N=2$ supersymmetric Teichm\"uller deformations or the $N=2$
super-quadratic differentials. The $N=2$ supersymmetric Teichm\"uller space
$\ct^{N=2}_{h,c}={\rm Ker}\cp_1^{\dg}$ can be naturally introduced in terms of
the original quantities by setting
$$\ct^{N=2}_{h,c}=\fracmm{\{E\du{M}{A},~\O_N;~\d(T)\}}{\{{\rm 2sDiff}_0(\X)
\otimes {\rm 2sWeyl}(\X)\otimes {\rm 2sL}\otimes {\rm 2sU(1)}\}}~,\eqno(4.14)$$
where the $N=2$ supergravity constraints '$T=0$' are supposed to be satisfied,
 and ${\rm 2sDiff}_0(\X)$ means the group of topologically trivial
diffeomorphisms  connected to the identity. Since any non-trivial
topology of the $N=2$ SRS is due to its 'body' (which is an ordinary Riemann
surface), the quotient ${\rm 2sDiff}(\X)/{\rm 2sDiff}_0(\X)$ is the {\it
ordinary} mapping class group ${\rm MCG}_h$, ${\rm MCG}_h={\rm Diff}(\X|)/{\rm
 Diff}_0(\X|)$.

The $N=2$ super-moduli space is defined by
$$\cm^{N=2}_{h,c}=\fracmm{\ct^{N=2}_{h,c}}{{\rm MCG}_h}~.\eqno(4.15)$$
Since the $N=2$ superconformal structure (4.9) on $N=2$ SRS is already
invariant with respect to the local symmetries (ii), (iii) and (iv), the
equivalent definition of the $N=2$ super-moduli space is given by
$$\cm^{N=2}_{h,c}=\fracmm{\{J\}}{{\rm 2sDiff}(\X)}~.\eqno(4.16)$$

In analysing the $N=2$ super-moduli space of $N=2$ SRS, it is sometimes useful
to consider the $N=2$ supergeometries characterised by a constant
supercurvature, $R=const$. The 'constant-curvature' $N=2$ supergeometries
comprise the globally defined slice with respect to the $N=2$ super-Weyl
transformations in the space of all $N=2$ supergeometries, when the $N=2$
super-Liouville equation (4.2) shows an action of these transformations along
the orbits. One gets, in particular, from eqs.~(4.14) and (4.15) that
$$\cm^{N=2}_{h,c}=\fracmm{\{E\du{M}{A},~\O_N;~\d(T)\}_{R=const}}{\{{\rm
2sDiff}(\X)\otimes {\rm 2sL}\otimes {\rm 2sU(1)}\}}~.\eqno(4.17)$$
The $N=2$ super-moduli space $\cm^{N=2}_{h,c}$ is a supermanifold, whose
tangent space is the space spanned by all $N=2$ supersymmetric Teichm\"uller
deformations ${\rm Ker}\cp_1^{\dg}$. Unfortunately, it is not clear for us at
the moment, how to get more insights into the complicated structure of the
$N=2$ super-moduli space, while keeping in use the $N=2$ superspace approach
and not going to the components.
\vglue.2in

\section{$N=2$ Super-Weyl Anomaly}

Given the action (4.5), the Green's functions or the propagators of the $N=2$
chiral scalar superfields $X$ and $\bar{X}$ satisfy the equations
$$\left(\begin{array}{cc} 0 & -\ha(\bar{\de}^2-4R) \\
 -\ha(\de^2-4\bar{R}) & 0 \end{array}\right)\left(\begin{array}{cc} 0 & G_{\rm
ca} \\ G_{\rm ac} & 0\end{array}\right)=\left(\begin{array}{cc}\d^4_{\rm c} & 0
\\ 0 & \d^4_{\rm a}\end{array}\right)~,\eqno(5.1)$$
where the chiral delta-functions have been introduced,
$$\eqalign{
\d^4_{\rm c} = & -\ha(\bar{\de}^2-4R) E^{-1}\d^6(z,z')~,\cr
\d^4_{\rm a} =  & -\ha(\de^2-4\bar{R}) E^{-1}\d^6(z,z')~,\cr}\eqno(5.2)$$
and
$$\d^6(z,z')\equiv\d^2(x,x')\d^2(\q-\q')\d^2(\bar{\q}-\bar{\q'})~.\eqno(5.3)$$

The integration over the $N=2$ matter fields in the partition function (4.7)
yields
$$\int \[d\bar{X}\]\[dX\]e^{-S_0}=e^{-W}~,\eqno(5.4)$$
where
$$W=-\ha{\rm sTr}\ln G +{\rm h.c.}~,\eqno(5.5)$$
and
$$ G\equiv G_{\rm ca}G_{\rm ac}~,\quad \bar{G}\equiv G_{\rm ac}G_{\rm ca}~.
\eqno(5.6)$$
The Green's functions $G_{\rm ac}$ and $G_{\rm ca}$ can now be written down in
the form
$$\eqalign{
G_{\rm ac}= & -\ha(\de^2-4\bar{R})G~,\cr
G_{\rm ca}= & -\ha(\bar{\de}^2-4R)\bar{G}~,\cr}\eqno(5.7)$$
where the new ones $G$ and $\bar{G}$ satisfy the equations
$$\eqalign{
\ch G\equiv & \fracmm{1}{4}(\bar{\de}^2-4R)(\de^2-4\bar{R})G=\d^4_{\rm c}~,
\cr
\bar{\ch}\bar{G}\equiv & \fracmm{1}{4}(\de^2-4\bar{R})(\bar{\de}^2-4R)
\bar{G}=\d^4_{\rm a}~.\cr}\eqno(5.8)$$
The $\ch$-operators can be thought of as the $N=2$ (chiral) covariant
scalar 'Laplacians' squared \cite{a83}, {\it viz.}
$$\eqalign{
\ch =\fracmm{1}{4}(\bar{\de}^2-4R)(\de^2-4\bar{R}) & =\fracmm{1}{4}\left[
\bar{\de}^2\de^2-4\bar{\de}^2\bar{R}-4R\de^2+16R\bar{R}\right]~,\cr
\bar{\ch}=\fracmm{1}{4}(\de^2-4\bar{R})(\bar{\de}^2-4R) & =
\fracmm{1}{4}\left[ \de^2\bar{\de}^2 -4\de^2 R-4\bar{R}\bar{\de}^2
+16\bar{R}R\right]~.\cr}\eqno(5.9)$$

The formal expression ${\rm sTr}\ln G=\ln{\rm sdet} G=-\ln{\rm sdet}\ch$
needs to be regularised,
and it has to be carried over the space orthogonal to the zero modes of $\ch$:
${\rm sdet}\to{\rm sdet}'$. The natural definition is \cite{dp88r,ma83}
$$\ln{\rm sdet}'(\ch + s)=-\int^{\infty}_{\ve}\fracmm{dt}{t}\,e^{-ts}{\rm
sTr}'\left[e^{-t\ch}\right]~,\eqno(5.10)$$
where the real UV cutoff $\ve$ and the complex parameter $s$ have been
introduced. The integral in eq.~(5.10) absolutely converges for sufficiently
large ${\rm Re}(s)$ and $\ve>0$. The definition can then be extended throughout
the complex $s$-plane by analytic continuation. The limit $s\to 0$ determines
the regularised superdeterminant we are interested in.

The infinitesimal variation
$$\d\ln{\rm sdet}'(\ch + s)=\int^{\infty}_{\ve}dt\,e^{-ts}{\rm sTr}'\left[\d
\ch e^{-t\ch}\right]\eqno(5.11)$$
under the $N=2$ super-Weyl transformations with the infinitesimal $N=2$
chiral superfield local parameter $\d\S$ can be explicitly computed since
eqs.~(2.21), (2.22) and (5.9). The finite $N=2$ super-Weyl transformation of
the $\ch$-operator takes the form
$$\hat{\ch}=e^{-2\S}e^{-2\bar{\S}}\left\{ \ch + 8(\bar{\de}^2-4R)\de^2\S
\right.$$
$$\left. - [(\bar{\de}^{\dt{\a}}\bar{\S})\bar{\de}_{\dt{\a}}
- 6(\bar{\de}^2\bar{\S}) + 12(\bar{\de}\bar{\S})^2](\de^2-4\bar{R} + 8\de^2\S)
\right\}~.\eqno(5.12)$$
Fortunately, in order to compute the $N=2$ super-Weyl anomaly whose local form
can be fixed up to a coefficient (see below) on the symmetry grounds, knowing
the infinitesimal chiral part of the {\it rigid} $N=2$ super-Weyl
transformation of $\ch$, $\d\ch = -2\ch\d\S +\ldots\,$, is enough.  Therefore,
 keeping only the relevant term, we find
$$\lim_{s\to 0}\d\ln{\rm sdet}'(\ch +s)=-2\lim_{s\to 0}\int^{\infty}_{\ve}dt\,
e^{-ts}{\rm sTr}'\left[\d\S \ch e^{-t\ch}\right]$$
$$=2\lim_{s\to 0}\int^{\infty}_{\ve}dt\,e^{-ts}\fracmm{\pa}{\pa t}\left(
{\rm sTr}'\left[\d\S e^{-t\ch}\right]\right)=-2{\rm sTr}'\left[
\d\S e^{-\ve\ch}\right]~.\eqno(5.13)$$
In the limit $\ve\to 0$ the expression on the right-hand side of eq.~(5.13) is
local, and it is entirely determined by the symmetry, locality and chirality
arguments  up to an overall constant, namely
$$\lim_{\ve\to 0}{\rm sTr}'\left[\d\S e^{-\ve\ch}\right] ={\rm const.}\int d^2x
d^2\q\,\ce R\d\S~.\eqno(5.14)$$
There is no $1/\ve$ (divergent) term due to world-sheet supersymmetry
\cite{dp88r,ma83}. The constant in eq.~(5.14) can be computed \footnote{In
fact, the actual form of the anomaly in eq.~(5.14) also follows from the
short-time expansion, as can be shown by a straightforward (tedious)
calculation.} via a short-time expansion of the $N=2$ chiral super-heat kernel
$U(z,z';t)$ to be defined as the solution to the equation
$$\left(\fracmm{\pa}{\pa t}+\ch\right)U(z,z';t)=\d(t)\d^4_{\rm c}(z,z')~.
\eqno(5.15)$$
Since we are actually interested in calculating the elements of $U$ on the
diagonal $z=z'$ for short times $t\sim \ve\to +0$, which is a {\it local}
problem, we use the $N=2$ (non-anomalous) local symmetries to render the $N=2$
 supergeometry to be superconformally flat, i.e. take the reference $N=2$
supergeometry ('with hats') to be flat. Moreover, since the anticipated
structure of the result in eq.~(5.14), we keep only those terms in $\ch$ which
are {\it linear} in $\S$ and have no more than {\it two} derivatives acting on
$\S$. \footnote{It can be shown that the other contributions vanish in the
limit $t\to 0$ \cite{dp88r,ma83}.} Finally, we temporarily omit the constant
scaling factor $e^{-2\S}$ in the expression for the $\ch$-operator, in order
to restore it at the end. After all that simplifications the remaining terms
of $\ch$ read:
$$\ch\to\ch_{\rm re}=\bar{D}^2D^2-C^{\dt{\a}\dt{\b}}(\bar{D}_{\dt{\b}}\bar{\S})
\bar{D}_{\dt{\a}}D^2 + 6(\bar{D}^2\bar{\S})D^2~,\eqno(5.16)$$
where we have used the conventional notation for the flat $N=2$ superspace
covariant derivatives:
$$\bar{D}^2=\ha C^{\dt{\a}\dt{\b}}\bar{D}_{\dt{\b}}\bar{D}_{\dt{\a}}~,\quad
D^2=\ha C^{\a\b}D_{\b}D_{\a}~,\eqno(5.17)$$
as well as the identities
$$C^{\dt{\a}\dt{\b}}\bar{D}_{\dt{\b}}(\g_3)\du{\dt{\a}}{\dt{\g}}
\bar{D}_{\dt{\g}}=C^{\a\b}(\g_3)\du{\b}{\g}D_{\g}D_{\a}=0~,$$
$$D^2\bar{D}^2D^2=\pa_a\pa^a D^2\equiv \D D^2~.\eqno(5.18)$$
The $\ch_{\rm re}$ can then be rewritten to the form $\ch_{\rm re}
=(\D -\cv_{\rm re})D^2$, with $\cv_{\rm re}$ to be considered as
a perturbation,
$$\cv_{\rm re}=C^{\dt{\a}\dt{\b}}(\bar{D}_{\dt{\b}}\bar{\S})\bar{D}_{\dt{\a}}
-6\bar{D}^2\bar{\S}~.\eqno(5.19)$$

The {\it flat} chiral $N=2$ supersymmetric heat kernel equation
$$\left(\fracmm{\pa}{\pa t}+\D\right)U(z,z';t)=\d^2(x,x')\d^2(\q-\q')\d(t)~
\eqno(5.20)$$
is solved by
$$U_0(z,z';t)=U_0(x,x';t)\d^2(\q-\q')~,\eqno(5.21)$$
where the usual flat space heat kernel
$$U_0(x,x';t)=\fracmm{1}{4\p t}e^{-\fracmm{(x-x')^2}{4t}}\vq(t)~,\eqno(5.22)$$
has been introduced. \footnote{The $\vq(t)$ denotes the conventional
step-function: $\vq(t)=0$ when $t<0$, and $\vq(t)=1$ when $t>0$.} The only
term we need to consider in the iterative solution for the $U(z,z';t)$ is
$$U_{\rm re}(z,z';t)=\int^t_0 dt'\int d^2x' d^2\q'\,U_0(z,z';t-t')\cv_{\rm re}
(z',\q')U_0(z,z';t')+O(t)$$
$$=\int^t_0 dt'\int d^2x'\,U_0(x,x';t-t')U_0(x,x';t')\int d^2\q'\,\d^2(\q,\q')
\cv_{\rm re}(z',\q')\d^2(\q,\q')+O(t)~.$$
The integral over $\q'$ in the last line of this equation contributes
$$\cv_{\rm re}(z',\q')\d^2(\q,\q')|_{\q'=0}=-6\bar{D}^2\bar{\S}~,$$
whereas the remaining integral over $t'$ gives the factor $(2\p)^{-1}$ in the
limit $t\to 0$. Putting it all together, we find
$$\lim_{t\to 0}U_{\rm re}(z,z;t)=-\fracmm{6}{2\p}\bar{D}^2\bar{\S}~,
\eqno(5.23a)$$
and, hence, after restoring the constant scaling by $\S$, the covariant form
of the solution reads
$$\lim_{t\to 0}U(z,z;t)=-\fracmm{3}{2\p}R~.\eqno(5.23b)$$
Given $d$ chiral scalar $N=2$ superfields, the anomalous contribution of
eq.~(5.23) should be multiplied by $d$. Therefore, the constant in eq.~(5.14)
is equal to $-3d/(2\p)$.

Eqs.~(5.5), (5.13), (5.14) and (5.23) imply for the anomalous part of the
induced  action $W$
$$\d W_{\rm anomalous}=\fracmm{3d}{2\p}\int d^2x d^2\q\,\ce R \d\S+{\rm h.c.}~,
\eqno(5.24)$$
whose integration yields
$$W_{\rm anomalous}=-\fracmm{3d}{2\p}\int d^2x d^2\q d^2\bar{\q}\,\hat{E}
\bar{\S}\S +\left\{\fracmm{3d}{2\p}\int d^2x d^2\q\,\hat{\ce}\hat{R}\S
+ {\rm h.c.} \right\}~,\eqno(5.25)$$
where all the $N=2$ superderivatives and supercurvatures are to be defined
with respect to the reference (e.g., of constant supercurvature) $N=2$
superspace geometry. The complete expression for the $W$  contains the
additional anomaly-free factor
$$\ha\ln\fracmm{{\rm sdet}'\hat{\ch}}{{\rm sdet}\VEV{\hat{\F}_I|\hat{\F}_J}
{\rm sdet}\VEV{\hat{\J}_I|\hat{\J}_J}}+{\rm h.c.}~,\eqno(5.26)$$
where the (orthonormal) basises $\{\F_I\}$ and $\{\J_I\}$ of the
finite-dimensional spaces ${\rm Ker}\,(\bar{\de}^2-4R)$ and ${\rm Ker}\,(\de^2
-4\bar{R})$, respectively, have been introduced.

The $N=2$ superfield supergravity measure yields, in its turn, the
contribution to the partition function (4.7),
$$\[dE\du{M}{A}\]\[d\O_N\]\d(T)=\left({\rm sdet}'\cp^{\dg}_1\cp_1\right)^{1/2}
\fracmm{1}{{\rm Vol}({\rm Ker}\cp_1)}\[d\S\]\[d\bar{\S}\]\[d'V_A\]\[dL\]
\[dM\]~,\eqno(5.27)$$
where some factors cancel after the appropriate normalisation of the partition
 function by
$$\cn^{-1}={\rm Vol}({\rm 2s Diff}_0)\times{\rm Vol}({\rm 2sL})\times
{\rm Vol}({\rm 2sU(1)})~.\eqno(5.28)$$
The $N=2$ superghosts associated with the non-trivial factor
$({\rm sdet}'\cp^{\dg}_1\cp_1)^{1/2}$ in the measure (5.27) contribute a local
factor to the $N=2$ super-Weyl anomaly, which should be similar to that in
eq.~(5.25), although we have not yet computed it. To match with the component
approach, it should give rise to a shift $d\to (d-2)$ in the total $N=2$
super-Weyl anomaly. The 'cosmological' term is also allowed to be added to the
final result, since it is consistent with the symmetries of the non-critical
$N=2$ fermionic string theory. Perhaps, it seems to be equally consistent to
set the 'cosmological' term to be zero, $\m_0=0$.

Finally, we arrive at the following expression for the $N=2$ super-Weyl
anomaly (with the anticipated ghost contribution included)
$$W_{\rm total,~anomalous}=-\fracmm{3(d-2)}{2\p}S_{N=2~{\rm Liouville}}
(\S,\bar{\S})~,\eqno(5.29)$$
where the $N=2$ super-Liouville action has been introduced,
$$S_{N=2~{\rm Liouville}}=\int d^2x d^2\q d^2\bar{\q}\,\hat{E}\bar{\S}\S
+\left\{\int d^2x d^2\q\,\left(\hat{\ce}\hat{R}\S+\m_0\ce\right) + {\rm h.c.}
\right\}\eqno(5.30)$$
In the $N=2$ superconformal gauge, it reduces to $(\hat{R}\equiv r^{-1}=const)$
$$S_{N=2~{\rm Liouville}}^{\rm g.-f.}=\int d^2x d^2\q d^2\bar{\q}\
\bar{\S}\S +\left\{\int d^2x d^2\q\,\left[r^{-1}\S + \m_0(e^{2\S}-1)\right]
+{\rm h.c.}\right\}\eqno(5.31)$$
The main results of this section are summarized by eqs.~(5.29) and (5.30).
The effective theory of the quantised $N=2$ supergravity in two dimensions is
given by the $N=2$ super-Liouville theory, as it should have been expected from
a consistency with the component approach. The current algebra of the $N=2$
super-Liouville theory in the $N=2$ superconformal gauge  was recently
discussed in refs.~\cite{s93a,gs93a}.
\vglue.2in

\section{Conclusion}

The results reported above are thought to be useful for a systematic covariant
quantisation of the $N=2$ fermionic string theory in $N=2$ superspace, which
is yet to be completed. A covariant derivation of the $N=2$ superfield BRST
operator, as well as an inspection of the BRST transformation properties of
the $N=2$ superfield path integral measure would be of particular interest.

There are several indications in the literature that both critical and
non-critical $N=2$ fermionic string theories are in fact {\it topological}
quantum field theories in the sense of ref.~\cite{w88}. Namely, the BRST
super-current algebra in these theories appears to be the particular $N=2$
supersymmetric extension of the topological conformal algebra \cite{w88} for
an arbitrary conformal matter \cite{gs93a}. As was noticed in ref.~\cite{gs93a}
for both  critical and non-critical $N=2$ fermionic string theories, the
vanishing of the ghost-number anomaly  associated with the total divergence
term in the full BRST anomaly  is crucial for the apparently topological
properties of these theories. There are also many reasons to believe that the
two-dimensional $N=2$ supergravity itself is a topological quantum field
theory \cite{k93,gs93a}.

An analysis of the zero modes of various $N=2$ super-differential operators
on the $N=2$ super-Riemannian surfaces turns out to be very involved in terms
of superfields, compared to the standard approach in components. This
currently appears to be the  major technical obstruction preventing an
efficient use of the covariant (not related with the $N=2$ superconformal
gauge) definition of the $N=2$ super-Riemannian
surfaces, which has been proposed in sect.~4.
 Nevertheless, nothing seems to be preventing, in principle, to formulate
$N=2$ super-analogue of the Riemann-Roch theorem which would be based on that
definition, and then explicitly construct the $N=2$ superfield measure at
non-vanishing values of genus $h\neq 0$ and Chern class $c\neq 0$.
It is currently under study.

One of the authors (S.V.K.) would like to thank S. J. Gates Jr. for useful
conversations.

\newpage

\section{Appendix: Notation}

The two-component (complex) world-sheet spinor is labelled as
$$\j^{\a}=(\j^+,\j^-)~,\eqno(A.1)$$
while its complex conjugate takes the form
$$\j^{\dt{\a}}=(\j^{\dt{+}},\j^{\dt{-}})~.\eqno(A.2)$$
The lower-case Greek letters are used to represent spinor indices, the
lower-case Latin letters are used for  tensor indices. The early
(both Greek and Latin) indices are referred to a tangent space, the middle
ones being referred to a base space. The capital letters are normally used
to denote $N=2$ superspace indices.

The two-dimensional Minkowski metric is
$$\h_{ab}={\rm diag} (-,+)~,\quad a,b=0,1~.\eqno(A.3)$$

The two-dimensional Dirac gamma matrices satisfy an algebra
$$\{\g^a,\g^b\}=2\h^{ab}~.\eqno(A.4)$$
Their explicit forms are
$$(\g^a)=(i\s_1,\s^2)~,\quad (\g_3)=\g^0\g^1=\s^3~.\eqno(A.5)$$

The Levi-Civita antisymmetric symbol in two dimensions is normalized by the
condition
$$ \ve^{01}=1~.\eqno(A.6)$$
The spinor metrics
$$C_{\a\b}=C_{\dt{\a}\dt{\b}}=\s^2 \eqno(A.7)$$
and their inverses are used to raise and lower the spinors indices:
$\j_{\a}=C_{\b\a}\j^{\b}$, $\j^{\a}=C^{\a\b}\j_{\b}$, and similarly
for the dotted indices.

The obvious identities take place
$$\ve^{ab}\ve_{cd}=\d_c^{[a}\d_d^{b]}~,$$
$$C^{\a\b}C_{\g\d}=\d_{\g}^{[\a}\d_{\d}^{\b]}~,$$
$$\g^a\g^b=\h^{ab}+\ve^{ab}\g_3~,$$
$$\j^2=\ha C_{\b\a}\j^{\a}\j^{\b}=\ha \j^{\a}\j_{\a}=i\j^+\j^-~.\eqno(A.8)$$

Given a general supermatrix
$$ M = \left( \begin{array}{cc} A\du{a}{b} &  B\du{a}{\b,\dt{\b}} \\
C\du{\a,\dt{\a}}{b} & D\du{\a,\dt{\a}}{\b,\dt{\b}}\end{array} \right)~,
\eqno(A.9)$$
its superdetermiant and supertrace are defined by
$${\rm sdet}\,M=\det\left(A-BD^{-1}C\right)\det{}^{-1}D~,$$
$${\rm str}\,M=\tr\,A - \tr\,D~.\eqno(A.10)$$

The delta-functions of the anticommuting coordinates $\q_{\a}$ and
$\bar{\q}_{\dt{\a}}$ are defined by
$$\d^2(\q_1 -\q_2)=(\q_1 -\q_2)^2~,\quad \d^2(\bar{\q}_1 -\bar{\q}_2)
=(\bar{\q}_1 -\bar{\q}_2)^2~,\eqno(A.11)$$
so that
$$\int d^2\q\,\d^2(\q)=\int d^2\bar{\q}\,\d^2(\bar{\q})=1~.\eqno(A.12)$$
\vglue.2in


\newpage

\begin{thebibliography}{99}

\bibitem{p81} A. M. Polyakov, \pl{103}{81}{207}, {\it ibid.} 211.
\bibitem{h79} P. Howe, J. Phys. {\bf A12} (1979) 393.
\bibitem{dp88r} E. D'Hoker and D. H. Phong, Rev.~Mod.~Phys. {\bf 60} (1988)
917.
\bibitem{bmg86} R. Brooks, F. Muhammad and S. J. Gates, Jr.,
\np{268}{86}{599}.
\bibitem{ft81} E. S. Fradkin and A. A. Tseytlin, \pl{106}{81}{63}.
\bibitem{bn86} P. Bouwknegt and P. van Nieuwenhuizen, \cqg{3}{86}{207}.
\bibitem{mm88} S. D. Mathur and S. Mukhi, \np{302}{88}{130}.
\bibitem{dhk90} J. Distler, Z. Hlousek and H. Kawai, \ijmp{5}{90}{391}.
\bibitem{abk90} L. Antoniadis, C. Bachas and C. Kounnas, \pl{242}{90}{185}.
\bibitem{hp87} P. Howe and G. Papadopoulos, \cqg{4}{87}{11}.
\bibitem{glo89} S.~J.~Gates Jr., L.~Lu and R.~Oerter, \pl{218}{89}{33}.
\bibitem{ghn89} S.~J.~Gates Jr., S.~J.~Hassoun and P. van Nieuwenhuizen,
\np{317}{89}{302}.
\bibitem{a90} A. Alnowaiser, \cqg{7}{90}{1033}.
\bibitem{a71m} M. Atiyah, Ann.~Sci.~Ec.~Norm.~Sup. {\bf 4} (1971) 47.
\bibitem{dp86m} L. Dabrowski and R. Percacci, \cmp{107}{86}{691}.
\bibitem{adp89} K. Aoki, E. D'Hoker and D. H. Phong, \np{342}{90}{149}.
\bibitem{ggrs} S.~J.~Gates Jr., M.~T.~Grisaru, M.~Rocek and W.~Siegel,
{\it Superspace or One Thousand and One Lessons in Supersymmetry},
Benjamin-Cummings, MA, 1983.
\bibitem{ghr84} S. G. Gates, Jr., C. M. Hull and M. Ro\v{c}ek,
\np{248}{84}{570}.
\bibitem{gsw} M. B. Green, J. H. Schwarz and E. Witten {\it Superstring
Theory}, Vol. I, Cambridge University Press, Cambridge, 1987.
\bibitem{p86u} J. Polchinski, \cmp{104}{86}{37}.
\bibitem{k93} S. V. Ketov, {\it Space-Time Supersymmetry of Extended Fermionic
Strings in $2+2$ Dimensions}, preprint Hannover ITP-UH-1/93 and DESY 93-050,
February 1993, to appear in ''Classical and Quantum Gravity''.
\bibitem{dw83} B. De Witt, {\it Supermanifolds}, Cambridge University Press,
Cambridge, 1983.
\bibitem{c87} I. D. Cohn, \np{284}{87}{349}.
\bibitem{m88} E. Melzer, \jmp{29}{88}{1555}.
\bibitem{a83} O. Alvarez, \np{216}{83}{125}.
\bibitem{ma83} E. Martinec, \pr{28}{83}{2604}.
\bibitem{s93a} W. A. Sabra, \np{392}{93}{385}.
\bibitem{gs93a} J. Gomis and H. Suzuki, \np{393}{93}{126}.
\bibitem{w88} E. Witten, \cmp{177}{88}{353}, \ibid{118}{88}{411}.

\end{thebibliography}
\end{document}